\documentclass[12pt]{article}
\usepackage{geometry}
\usepackage{times}
\usepackage[hidelinks]{hyperref}       
\usepackage{booktabs, threeparttable, siunitx}       
\usepackage{amsmath, amssymb, amsthm, bbm, mathtools, mathrsfs}       
\usepackage{nicefrac}       
\usepackage{xcolor}
\usepackage{enumitem}
\usepackage{multirow}
\usepackage{natbib}
\usepackage{graphicx}
\usepackage{setspace}
\usepackage{endfloat}

\usepackage{etoolbox}
\robustify\bfseries

\makeatletter
\newcommand{\distas}[1]{\mathbin{\overset{#1}{\kern\z@\sim}}}%
\makeatother
\newcommand{\E}{\mathbb{E}}
\newcommand{\prob}{\mathbb{P}}

\newcommand{\var}{\textnormal{Var}}

\newcommand\independent{\protect\mathpalette{\protect\independenT}{\perp}}
\def\independenT#1#2{\mathrel{\rlap{$#1#2$}\mkern2mu{#1#2}}}
\definecolor{amber}{rgb}{1.0, 0.49, 0.0}

\newtheorem{proposition}{Proposition}
\newtheorem{assumption}{Assumption}
\newtheorem{lemma}{Lemma}
\newtheorem{theorem}{Theorem}

\theoremstyle{remark}

\title{Entropy Balancing for Causal Generalization with Target Sample Summary Information}

\author{Rui Chen$^{1}$,
	Guanhua Chen$^{2}$\thanks{gchen25@wisc.edu},
    Menggang Yu$^{2}$\thanks{meyu@biostat.wisc.edu} \bigskip \\
	$^{1}$Department of Statistics,\\ University of Wisconsin-Madison\\ [8pt]
	$^{2}$Department of Biostatistics and Medical Informatics,\\ University of Wisconsin-Madison \\ [8pt]}

\begin{document}
	
	\renewcommand{\abstractname}{Summary}
	\singlespacing
	\maketitle
	\thispagestyle{empty}
	\setstretch{1.85}
	
	\begin{abstract}
	In this paper, we focus on estimating the average treatment effect (ATE) of a target population when individual-level data from a source population and summary-level data (e.g., first or second moments of certain covariates) from the target population are available. In the presence of heterogeneous treatment effect, the ATE of the target population can be different from that of the source population when distributions of treatment effect modifiers are dissimilar in these two populations, a phenomenon also known as covariate shift. Many methods have been developed to adjust for covariate shift, but most require individual covariates from a representative target sample. We develop a weighting approach based on summary-level information from the target sample to adjust for possible covariate shift in effect modifiers. In particular, weights of the treated and control groups within a source sample are calibrated by the summary-level information of the target sample. Our approach also seeks additional covariate balance between the treated and control groups in the source sample. We study the asymptotic behavior of the corresponding weighted estimator for the target population ATE under a wide range of conditions. The theoretical implications are confirmed in simulation studies and a real data application.
	
	\medskip
	
	{{\it Keywords:} Average Treatment Effect, Causal Generalization, Entropy Balancing Weights, Summary-Level Data}
	\end{abstract}
	\thispagestyle{empty} 
	\newpage
	\setcounter{page}{1}
	

\setstretch{1.85}
\section{Introduction}

It is often of interest to apply causal findings from a medical study in one population (a source population) to another (a target population) based on observed characteristics  \citep{colnet2020causal,degtiar2021review}. This problem is termed generalizability \citep{Cole2010,tipton2013improving,buchanan2018generalizing}, external validity \citep{Rothwell2005}, or transportability \citep{Rudolph2017,dahabreh2019extending}. It is well-known that such generalization may be problematic when the causal effect depends on certain covariates (also known as effect modifiers) and there is covariate shift in the effect modifiers \citep{sugiyama2007covariate}.  Covariate shift refers to a shift or a difference in the distribution of a covariate between the source and target populations. 

For example, average treatment effect (ATE) from a properly planned and conducted randomized controlled trial (RCT) may not be generalizable when the treatment effect is modified by   covariates and these covariates can have different distributions between the study population and the target population for generalization. In other words, we could obtain an unbiased estimate of ATE for the trial population, but the estimate may not equal the ATE of the target population if the study participants do not represent the target population well with respect to the treatment effect modifiers.

In the past decade, a common setup for such generalization of causal findings is based on the scenario that individual-level covariates are available for the target population \citep{Cole2010,tipton2013improving,Rudolph2017,buchanan2018generalizing,dahabreh2020extending,lu2021you}. Under this scenario, most existing methods rely on modeling trial or source sample participation probability, which reflects similarity between subjects in the two populations. The probability is modeled using individual-level data and the estimated probability is then used in the subsequent analysis for reweighting \citep{Cole2010,buchanan2018generalizing} or post-stratification \citep{Cole2010}. 

Some existing methods also incorporate outcome modeling to improve estimation efficiency \citep{Rudolph2017,dahabreh2020extending,yang2020doubly}. Similarly,  fitting outcome models also needs individual-level target sample data for generalization.

Nevertheless, detailed individual-level information is not always available from a target sample due to many practical reasons such as restricted data sharing, storage limitation, and privacy concerns \citep{degtiar2021review}. In contrast, summary-level information of the target sample is more accessible. Such information can be collected from population-based census data, disease registries and health care databases, and published literature.  \cite{josey2020calibration} investigated causal generalization based on summary-level information of a target sample.

In this paper, we extend \cite{josey2020calibration} to a weighting strategy that can better utilize individual-level covariates from a source sample for  causal generalization 
 \citep{hartman2015sample,westreich2017transportability}.  In particular, we  apply an entropy balancing weighting framework \citep{hainmueller2012entropy} to the summary-level information of a target sample, as well as to the treatment-control differences (based on individual-level covariates) of the source sample.  Our theoretical results show that the proposed method can not only achieve consistent estimation under significantly broader situations, but also result in higher estimation efficiency.

To balance individual-level treatment-control covariate differences in the source sample, more flexible strategies can be used. For example, for a covariate whose distribution is skewed and heavy-tailed,  balancing can be based on moments of its transformation, instead of its original value, especially when the transformed distribution is approximately normal. The intuition is that normal distribution is completely determined by its first two moments; hence balancing on these moments is equivalent to balancing the whole distribution.

This paper is organized as follows. After reviewing related entropy balancing methods in Sections \ref{sec:framework} and \ref{sec:method_review}, we develop our weighting approach  in Section \ref{sec:ebal_xbal}. In Section \ref{sec:theory}, we characterize the theoretical properties of the proposed method. Sections \ref{sec:simu} and \ref{sec:data} compare our method to other weighting methods by simulation studies and a real data application, respectively. We conclude the paper with a discussion in Section \ref{sec:discussion}.

\section{Notation and Framework}
\label{sec:framework}

Suppose we have collected data in a representative sample  $\mathcal{S}$ with $n_s$ subjects, $\{(X_i, A_i, Y_i): i \in \mathcal{S}\}$, from a source population. Here $X_i \in \mathcal{X} \subset \mathbb{R}^p$ is a vector of pre-treatment covariates which contain confounding factors and treatment effect modifiers, $A_i \in\{0, 1\}$ is a binary treatment indicator, and $Y_i$ is the outcome of interest. 

Now suppose we have a representative sample $\mathcal{T}$ with $n_t$ subjects from a target population.  However the individual-level data, $\{(X_i, A_i, Y_i): i \in \mathcal{T}\}$, is not observed. Actually the treatment and outcome data, $\{(A_i, Y_i): i \in \mathcal{T}\}$, does not even need to be defined. In addition, instead of observing the individual-level covariates, $\{X_i: i \in \mathcal{T}\}$, we only have access to first moments based on a set of linearly independent covariate functions $h_k: \mathcal{X} \rightarrow \mathbb{R}, ~ k = 1,\dots,K_h$.  Namely, we only have the following information from the target sample:
\begin{equation}
	\bar{h}_{k, \mathcal{T}} \equiv \frac{1}{n_t} \sum_{i \in \mathcal{T}} h_k(X_i), \ k = 1,\dots,K_h. \label{summary info from target}
\end{equation}

Frequently, ${h_k}$ is defined on one or two covariates, instead of on the full covariate vector. For continuous covariates, $h_k$ can be an identity function defined on some component of $X$ and then $\bar{h}_{k, \mathcal{T}}$ is the mean of this component. Alternatively $h_k$ can be a polynomial function (e.g. with degree 2)  and then $\bar{h}_{k, \mathcal{T}}$ corresponds to the second moment or variance of this component. For discrete covariates, $h_k$ can be an indicator function  to count the number of subjects in a particular category of a discrete variable.

The number of summary functions $K_h$ can be less than the number of covariates $p$. Such a situation can arise when aggregate-level information of only certain covariates is available for the target sample, while individual-level information of more covariates is collected in the source sample. The main contribution of our method is to fully utilize the extra individual-level covariate information of the source sample to gain efficiency in estimating or generalizing the ATE for the target population, which we will describe in detail in Section \ref{sec:ebal_xbal}.

We use the potential outcome framework \citep{rubin1974estimating,rosenbaum1983central} to formulate the causal problem. Under the Stable Unit Treatment Value Assumption (SUTVA), which posits no interference between different subjects and no hidden variation of treatments, each subject $i$ has two potential outcomes $Y_i(0)$ and $Y_i(1)$, the values of the outcome that would be observed if $i$ were to receive control or treatment, respectively. Then the observed outcome in the source sample is $Y_i = Y_i(A_i)$. For subjects in the target sample, neither of the potential outcomes is observed.

We associate each subject, either in the source or target sample, with a ``full'' random vector $(X_i , S_i , A_i , Y_i(0), Y_i(1))$, where $S_i$ is a population indicator such that $S_i = 1$ for $i \in \mathcal{S}$ and $S_i = 0$ for $i \in \mathcal{T}$. For $i \in \mathcal{T}$, $A_i$ can take arbitrary value and will not affect the following analysis. The total sample size is $n = n_s + n_t$. These $n$ random variates across $i$ are assumed to be i.i.d. draws from a joint distribution of $(X, S, A, Y(0), Y(1))$. All the probability calculations and expectations below are taken with respect to this distribution. Specifically, the ATE of the target population can be expressed as
\begin{equation*}
	\tau^* = \E\left[ Y(1) - Y(0) \mid S = 0 \right]
,\end{equation*}
which is the estimand of interest in this paper.

We assume that the treatment assignment mechanism in the source sample is determined by a \textit{propensity score} $\pi(x) = \prob (A = 1 \mid X = x, S=1)$ \citep{rosenbaum1983central}, which is potentially unknown. We further denote $\rho(x) = \prob(S=1 \mid X=x)$ and refer to this as \textit{participation probability} \citep{dahabreh2020extending}. In addition to the SUTVA, we impose the following identifiability assumptions throughout the paper.

\begin{assumption}
\label{assp:unconf}
    (Unconfoundedness of treatment assignment)
    In the source population, $(Y(0), Y(1))$ are conditionally independent of $A$ given $X$:
    $(Y(0), Y(1)) \independent A \mid X, S=1$.
\end{assumption}
\begin{assumption}
\label{assp:ovlptrt}
    (Positivity of propensity score)
    The propensity score of the source population is bounded away from 0 and 1: for some $c > 0$, $c \le \pi(X) \le 1 - c$ almost surely. 
\end{assumption}
\begin{assumption}
\label{assp:exchange}
    (Mean exchangability across populations)
    The conditional mean of the potential outcomes given the covariates are equal between the two populations: $\E [Y(a) \mid X, S=1] = \E [Y(a) \mid X, S=0]$ almost surely for $a \in \{0, 1\}$.
\end{assumption}
\begin{assumption}
\label{assp:ovlppop}
    (Positivity of participation probability)
    The participation probability is bounded away from 0: $\rho(X) > c$ almost surely for some $c > 0$. 
\end{assumption}

The first two are common assumptions in causal inference, and together with the SUTVA enable identification of causal quantities with respect to the source population from the observed data. The last two assumptions are adopted from \citet{Rudolph2017} and \citet{dahabreh2020extending} and allow us to generalize causal estimates to the target population \citep{colnet2020causal}. 

In what follows, we use $\mathcal{S}_0$ to denote the index set of the subjects in the control group of the source sample, namely, $\mathcal{S}_0 = \{i: S_i = 1, A_i = 0\}$; $\mathcal{S}_1$ is similarly defined for the treated group. We assume the potential outcomes have finite second moments given the covariates, and denote the conditional mean and variance of the potential outcomes in the source population by 
\begin{align*}
	\mu_{a}(x) &= \E\{Y(a)\mid X=x, S=1\},\\
	\sigma^2_{a}(x) &= \var\{ Y(a)\mid X=x, S=1\}
.\end{align*}
Under Assumption \ref{assp:exchange}, $\mu_a(x) = \E\{Y(a)\mid X=x, S=0\} = \E\{Y(a)\mid X=x\}$. The conditional average treatment effect (CATE) function is defined as $\tau(x) = \mu_1(x) - \mu_0(x)$.

\section{Review of existing approaches} 
\label{sec:method_review}

Under Assumptions 1-4, we can express $\tau^*$ in terms of the observable:
\begin{align}
\tau^*
&=
\E\left[ 
A w^*(1, X) Y - (1-A) w^*(0, X) Y \mid S = 1
\right]
\end{align}
where the weighting function $w^*(a, x)$ is given by
\begin{equation*}
w^*(a, x) = 
\left\{ \frac{a}{\pi(x)} + \frac{1-a}{1-\pi(x)} \right\}
\frac{\E(S)(1 - \rho(x))}{(1-\E(S)) \rho(x)} 
.\end{equation*}
Therefore given a set of weights on the source population data $\{w_i: i \in \mathcal{S}\}$, a possible estimator can take the following form:
\begin{equation}
	\label{eq:tau_hat_w}
	\hat{\tau}_w = 
	\frac{1}{n_s} \sum_{i \in \mathcal{S}_1} w_i Y_i
	-
	\frac{1}{n_s} \sum_{i \in \mathcal{S}_0} w_i Y_i
.\end{equation}

If there were access to the individual-level data of the target sample, $w^*(a, x)$ could be estimated by modeling $\pi(x)$ and $\rho(x)$. Under the current setting, however, usual modeling of $\rho(x)$ via maximum likelihood or loss minimization becomes infeasible. %

The idea is to use the target summary information \eqref {summary info from target} to calibrate the weights.

This calibration idea for weighted estimators has been extensively studied for causal inference \citep{hainmueller2012entropy, imai2014covariate, chan2016globally}. In the setting of no target population, \citet{hainmueller2012entropy} proposed to estimate average treatment effect on the treated (ATT) by constructing the weights for the control units as the solution to the following optimization problem:
\begin{equation}
\label{eq:ebal_ATT}
\begin{aligned}
\min_{w \succeq 0} \quad
& \sum_{i \in \mathcal{S}_0} w_i \log w_i
\\
\text{subject to} \quad
& \frac{1}{n_s} \sum_{i \in \mathcal{S}_0} w_i  X_i = \frac{1}{|\mathcal{S}_1|} \sum_{i \in \mathcal{S}_1} X_i \ \ \text{and }\  \frac{1}{n_s} \sum_{i \in \mathcal{S}_0} w_i = 1.
\end{aligned}
\end{equation}
Here $\succeq$ is the element-wise $\ge$ function for a vector. The balancing constraints equalize the sample means of the covariates between the treated group and the weighted control group; in the meantime, the optimization objective keeps the dispersion metric, the opposite of entropy, to a minimum level so that the weights are as close as uniform as possible.

\citet{zhao2016entropy} investigated theoretical properties of the weights given by \eqref{eq:ebal_ATT} and the corresponding ATT estimator. They showed the estimator enjoys the so-called doubly robustness property: if either the means of the potential outcomes are linear in the covariate functions being balanced, or the logit of the propensity score is linear in these functions, then the estimation is consistent. Further, if both conditions hold, then the semiparametric efficiency bound in \citet{hahn1998role} is achieved.

\citet{dong2020integrative} adapted the entropy balancing weights approach for generalizing ATE estimation from an RCT to a given target population. Since in an RCT the covariates of the treated and control groups are well-balanced by design, they did not distinguish these two groups in their weighting strategy and proposed to construct weights on the entire source sample $\{q_i: i \in \mathcal{S}\}$ by
\begin{equation}
    \label{eq:ebal_dong}
    \begin{aligned}
        \min_{q \succeq 0} \quad
        & \sum_{i \in \mathcal{S}} q_i \log q_i
        \\
        \text{subject to} \quad
		& \frac{1}{n_s} \sum_{i \in \mathcal{S}} q_i h_k(X_i) = \bar{h}_{k, \mathcal{T}}, ~ k = 0, \dots, K_h.
    \end{aligned}
\end{equation}
To normalize the weights so that $\sum_{i \in \mathcal{S}_0} q_i /{n_s} = 1$, we have included $h_0(x)=1$ as one of the constraints.

So the weights calibrate the sample averages of the covariate functions on the source sample to those on the target sample. By a duality argument, they showed that the solution to \eqref{eq:ebal_dong} admits the form of exponential tilting: $\hat{q}_i = \exp\{\sum_{k=0}^{K_h} \beta_k h_k(X_i)\}$ for some $\beta_k, k=0, \ldots, K_h$ such that the balancing constraints are satisfied. These weights take the same form as the exponential tilting adjustment method considered in \citet{signorovitch2010comparative}. In fact, the entropy balancing method based on \eqref{eq:ebal_dong} is equivalent to calibrating the covariate shift with an exponential tilting because the estimating equations that \citet{signorovitch2010comparative} utilized to estimate their exponential tilting parameters are exactly the balancing constraints in \eqref{eq:ebal_dong}. 
Using an application of generalizing the ATE of Rosuvastatin from an RCT to a target population, \citet{hong2019comparison} showed that the exponential tilting adjustment approach yielded close estimates to the one using individual-level target sample data for covariate shift adjustment.

\citet{josey2020calibration} then extended the entropy balancing weights approach to the setting when the source sample is from observational studies. To alleviate possible covariate imbalance or confounding between the treated and control groups in observational studies, they proposed a two-step procedure to adjust for covariate shift and confounding separately. They first computed $\{\hat{q}_i\}$ by solving \eqref{eq:ebal_dong}, and then used a subsequent step to further adjust for the treatment-control imbalance:
\begin{equation}
    \label{eq:ebal_josey}
    \begin{aligned}
        \min_{w \succeq 0} \quad
        & \sum_{i \in \mathcal{S}} w_i \log \left( \frac{w_i}{\hat{q}_i} \right) 
        \\
        \text{subject to} \quad
		& \frac{1}{n_s} \sum_{i \in \mathcal{S}_1} w_i h_k(X_i) = \frac{1}{n_s} \sum_{i \in \mathcal{S}} \hat{q}_i h_k(X_i), ~ k = 0, \dots, K_h\\
		& \frac{1}{n_s} \sum_{i \in \mathcal{S}_0} w_i h_k(X_i) = \frac{1}{n_s} \sum_{i \in \mathcal{S}} \hat{q}_i h_k(X_i), ~ k = 0, \dots, K_h.
    \end{aligned}
\end{equation}
So the constraints in \eqref{eq:ebal_josey} separately calibrate the treated and control groups to the same weighted source sample. Such balancing formulation bears similarity to ``template matching" \citep{silber2014template, bennett2020building}. The resulting weights can also be written as exponential tilting: $\hat{w}_i = \hat{q}_i \exp\{\sum_{k=0}^{K_h} \alpha_{ak} h_k(X_i)\}$, $a \in \{0, 1\}$. Both \citet{dong2020integrative} and \citet{josey2020calibration} established double robustness properties of the weighting estimators similar to \citet{zhao2016entropy}.

Alternatively, one could also use modeling-based weighting method to fully utilizes the individual-level source sample. Specifically, we can use the source sample to estimate the propensity score first, and then set the weights as $\hat{q}_i / \hat{\pi}(X_i)$ for the treated units and $\hat{q}_i / (1 - \hat{\pi}(X_i))$ for the control units. Since \eqref{eq:ebal_dong} is equivalent to modeling $\rho(x) / (1-\rho(x))$ with exponential tilting, the weights constructed in this way constitute estimates of $w^*(a, x)$. However, this approach is vulnerable to estimation error in the estimated propensity score and can result in extreme weights as one may encounter in the conventional inverse propensity score weights \citep{kang2007demystifying,chattopadhyay2020balancing}. In Section \ref{sec:simu}, we systematically compare this method with our proposed method. Numerical results suggest that our entropy balancing weighting approach leads to a more favorable performance.

\section{Method}
\label{sec:ebal_xbal}

We first show that the weights produced by the two-step procedure of \citet{josey2020calibration} can be consolidated into a one-step procedure. The simplified procedure simply takes $\hat{q}_i$ as 1 in \eqref{eq:ebal_josey}. Note that in this case, the right-hand sides of the constraints in \eqref{eq:ebal_josey} become $\bar{h}_{k, \mathcal{T}}$. Hence, the simplified procedure computes the weights by
\begin{equation}
    \label{eq:ebal_only_H}
    \begin{aligned}
        \min_{w \succeq 0} \quad
        & \sum_{i \in \mathcal{S}} w_i \log w_i
        \\
        \text{subject to} \quad
		& \frac{1}{n_s} \sum_{i \in \mathcal{S}_1} w_i h_k(X_i) = \bar{h}_{k, \mathcal{T}}, ~ k = 0, .\dots, K_h,
        \\& \frac{1}{n_s} \sum_{i \in \mathcal{S}_0} w_i h_k(X_i) = \bar{h}_{k, \mathcal{T}}, ~ k = 0, .\dots, K_h.
    \end{aligned}
\end{equation}

The equivalence between the two-step procedure of \citet{josey2020calibration} and \eqref{eq:ebal_only_H} is based on the dual representations of $\hat{q}_i$ and $\hat{w}_i$ in \eqref{eq:ebal_josey}. To see this, first note that the solutions from \eqref{eq:ebal_josey} take the exponential tilting form, $\exp\{\sum_k (\beta_k + \alpha_{ak}) h_k(X_i)\}$ for $i \in \mathcal{S}_a$ and that the weights given by \eqref{eq:ebal_only_H} also take the exponential tilting form $\exp\{\sum_k \lambda_{ak} h_k(X_i)\}$. So these two sets of weights have the same parametric form. Further, they satisfy the same set of constraints. When $\{h_k\}$ is linearly independent, we must have $\lambda_{ak} = \beta_k + \alpha_{ak}$. Therefore, \eqref{eq:ebal_only_H} produces exactly the same weights as \eqref{eq:ebal_josey} without first computing $\hat{q}_i$.

From \eqref{eq:ebal_only_H} we see that \citet{josey2020calibration} try to overcome covariate imbalance between the treatment and control groups in the source population by calibrating the covariate functions $h_k: \mathcal{X} \rightarrow \mathbb{R}, ~ k = 1,\dots,K_h$. However, these functions are made available to provide summary information from the target sample. They may not be flexible and sufficient for overcoming confounding of treatment assignment in the source sample.  

Therefore we propose to add further balancing requirements between the treated and control groups in the source sample to \eqref{eq:ebal_only_H} by utilizing an additional set of functions $\{g_k, ~k=1, \dots, K_g\}$. This gives rise to the following:
\begin{equation}
    \label{eq:ebal_HG}
    \begin{aligned}
        \min_{w \succeq 0} \quad
        & \sum_{i \in \mathcal{S}} w_i \log w_i
        \\
        \text{subject to} \quad
		& \frac{1}{n_s} \sum_{i \in \mathcal{S}_1} w_i h_k(X_i) = \bar{h}_{k, \mathcal{T}}, ~ k = 0, \dots, K_h
        \\
		& \frac{1}{n_s} \sum_{i \in \mathcal{S}_0} w_i h_k(X_i) = \bar{h}_{k, \mathcal{T}}, ~ k = 0, \dots, K_h
        \\
		& \frac{1}{n_s} \sum_{i \in \mathcal{S}_1} w_i g_k(X_i) = \frac{1}{n_s} \sum_{i \in \mathcal{S}_0} w_i g_k(X_i), ~ k = 1, \dots, K_g.
    \end{aligned}
\end{equation}
Without loss of generality, we assume the union set of covariate functions $\{h_k\} \cup \{g_k\}$ is linearly independent. To choose possible $g_k$, one can use those covariate functions that are deemed important for confounding but not covered in $\{\bar{h}_{k, \mathcal{T}}\}$. Our theoretic results in Section \ref{sec:theory} will give more guidance on the selection of $\{g_k\}$. It is worth noticing that \eqref{eq:ebal_only_H} can be viewed as a special case of \eqref{eq:ebal_HG} with $K_g = 0$.

To better understand the role of the weights given by \eqref{eq:ebal_HG} and facilitate theoretical investigation, the following proposition characterizes the dual problem. The proof of this proposition as well as other results in Section \ref{sec:theory} are relegated to the Web Appendix.
\begin{proposition} 
	\label{thm:solution_form}
	Let ${H} = (h_0, h_1, \dots, h_{K_h})$ and $G = (g_1, \dots, g_{K_g})$. The solution of \eqref{eq:ebal_HG} takes the following form:
    \begin{equation*}
		\hat{w}_i = \begin{cases}
			\exp\{ \hat{\lambda}_1^\mathsf{T}H(X_i) + \hat{\gamma}^\mathsf{T}G(X_i) \}, & i \in \mathcal{S}_1, \vspace{3pt}\\
			\exp\{ \hat{\lambda}_0^\mathsf{T}H(X_i) - \hat{\gamma}^\mathsf{T}G(X_i) \}, & i \in \mathcal{S}_0,
		\end{cases}
	\end{equation*}
   \vspace{-0.25in}
	where $(\hat{\lambda}_1, \hat{\lambda}_0, \hat{\gamma}) \in \mathbb{R}^{K_h+1} \times \mathbb{R}^{K_h+1} \times \mathbb{R}^{K_g}$ is the solution to the dual problem:
	\begin{equation}
		\label{eq:dual_problem_obj}
		\begin{aligned}
		\min_{\lambda_1, \lambda_0, \gamma} \quad
		&\frac{1}{n_s} \sum_{i \in \mathcal{S}_1} \exp\left\{ \lambda_1^\mathsf{T}H(X_i) + \gamma^\mathsf{T}G(X_i) \right\}
		\\&+
		\frac{1}{n_s} \sum_{i \in \mathcal{S}_0} \exp\left\{ \lambda_0^\mathsf{T}H(X_i) - \gamma^\mathsf{T}G(X_i) \right\} - 
		(\lambda_1 + \lambda_0)^\mathsf{T}\bar{H}_\mathcal{T}
		,\end{aligned}
	\end{equation}
	where $\bar{H}_\mathcal{T} = (\bar{h}_{0, \mathcal{T}}, \dots, \bar{h}_{K_h, \mathcal{T}})$.
\end{proposition}

Similar to the previous entropy balancing approaches, the weights given by \eqref{eq:ebal_HG} also  admit the form of exponential tilting, which is parameterized by a dual vector of length $2K_h + 2 + K_g$, the number of balancing constraints.

Proposition \ref{thm:solution_form} also reveals an efficient way to solve \eqref{eq:ebal_HG}. Since the dual problem is an unconstrained convex optimization with a closed-form derivative, it can be efficiently solved by common convex optimization algorithms such as the Newton-Raphson method. We note that the first-order optimality conditions of the dual problem \eqref{eq:dual_problem_obj} are exactly the balancing constraints in \eqref{eq:ebal_HG}. To ensure the solution feasibility of \eqref{eq:ebal_HG}, we need the number of constraints to be fixed, and Assumptions 2 and 4 to hold (i.e., there is sufficient overlap between treated and control groups within the source sample and between source and target samples). When the source sample size goes to infinity, the probability of the solution of \eqref{eq:ebal_HG} exists goes to one. Such a guarantee is a direct application of Proposition 1 of \citet{zhao2016entropy} for which the proof is included in the Web Appendix. The finite sample analysis of the solution feasibility is nontrivial and out of scope of this study.

Compared to the weights given by \eqref{eq:ebal_only_H}, the extended version \eqref{eq:ebal_HG} leads to better versatility as it allows the weights to depend on $\{g_k\}$, which can be a much broader set of covariate functions than $\{h_k\}$. However, unlike the association between the weights and $\{h_k\}$ where coefficients on $\{h_k\}$ for the treated group can be completely different from those for the control group, the association between the weights and $\{g_k\}$ is governed by a special structure: the coefficients for the two groups must be opposite of each other.

\section{Theoretical properties}
\label{sec:theory}

In this section, we study the theoretical properties of $\hat{\tau}_{\hat{w}}$ using weights given by the extended entropy balancing method \eqref{eq:ebal_HG}. Throughout we assume some mild regularity conditions. We first introduce the following lemma to characterize convergence of the weights.

\begin{lemma}
    \label{lem:lambda_gamma_limit_pi}
    Suppose $\log\{{\pi(x)}/{(1-\pi(x))} \} = \lambda_\pi^\mathsf{T}H(x) + \gamma_\pi^\mathsf{T}G(x)$ for some $\lambda_\pi, \gamma_\pi$. Let $(\hat{\lambda}_1, \hat{\lambda}_0, \hat{\gamma})$ be the dual parameters in Proposition \ref{thm:solution_form}. Then $(\hat{\lambda}_1, \hat{\lambda}_0, \hat{\gamma}) \overset{p}{\longrightarrow} (\lambda_0^* - \lambda_\pi, \lambda_0^*, -\gamma_\pi/2)$
    for some $\lambda_0^*$ satisfying $\E[\, \tilde{r}(X)H(X) \mid S=1\, ] = \E[H(X) \mid S=0]$ where
    \begin{equation}
	\label{eq:tilde_r}
        \tilde{r}(x) = \frac{ \exp\left\{ \lambda_0^{*\mathsf{T}}H(x) + \gamma_\pi^\mathsf{T}G(x)/2 \right\} }{ 1 + \exp\left\{ \lambda_\pi^\mathsf{T}H(x) + \gamma_\pi^\mathsf{T}G(x) \right\}}
    .\end{equation}
\end{lemma}

It follows from Proposition \ref{thm:solution_form} and Lemma \ref{lem:lambda_gamma_limit_pi} that the probability limits of the weights for the treated units can be obtained by the following function:
\begin{equation*}
	\exp\{ (\lambda_0^* - \lambda_\pi)^\mathsf{T} H(x) -\gamma_\pi^\mathsf{T} G(x) / 2 \}
	=
	\frac{1 + \exp\{\lambda_\pi^\mathsf{T}H(x) + \gamma_\pi^\mathsf{T}G(x)\}} {\exp\{\lambda_\pi^\mathsf{T}H(x) + \gamma_\pi^\mathsf{T}G(x)\}} \tilde{r}(x)
	= \frac{\tilde{r}(x)}{\pi(x)}
.\end{equation*}
Similarly, the limits of weights for the control units can be obtained from ${\tilde{r}(x)}/{(1-\pi(x))}$. In other words, when the propensity score follows the usual logistic regression w.r.t. the covariate functions $(H, G)$, the weights converge to the inverse of the propensity score multiplied by $\tilde{r}(x)$. Note that $\tilde{r}(x)$ is a positive function and satisfies $\E(\tilde{r}(X) \mid S=1) = \E(\tilde{r}(X) h_0(X) \mid S=1) = 1$, so $\tilde{r}(x)$ defines a density ratio over the source population. This means the weights asymptotically calibrate the observed source population data to a hypothetical population whose density ratio against the source population is $\tilde{r}(x)$. 

However, this hypothetical population is not necessarily the target population. Since the only available information from the target sample is the sample means on $H$, the covariate distribution of the target population is unidentifiable in general. Intuitively, our weighting approach essentially utilizes a parametric family of distributions to match these sample means. With this insight, the following theorem identifies conditions under which the resulting weighting estimator is consistent for $\tau^*$.

\begin{theorem}[Consistency]
    \label{thm:consistency_ebal_HG}
	Suppose $\hat{w}$ is the solution of \eqref{eq:ebal_HG}. If either of Conditions (a)-(c) below holds, $\hat{\tau}_{\hat{w}}$ is a consistent estimator of $\tau^*$:
    \begin{enumerate}[label=\textnormal{Condition (\alph*).}, itemsep=0pt, leftmargin=8em]
        \item $\mu_a(x) \in \textnormal{Span}\{H(x)\}$, $a = 0, 1$.
        \item $\log\{ {\pi(x)}/{(1-\pi(x))} \} \in \textnormal{Span}(H(x), G(x))$, and for some $\lambda$ the density ratio between the target population and the source population can be written as $\{ 1 + e^{ \lambda_\pi^\mathsf{T}H(x) + \gamma_\pi^\mathsf{T}G(x) }\}^{-1}{ e^{ \lambda^\mathsf{T}H(x) + \gamma_\pi^\mathsf{T}G(x)/2 } },$ where $\lambda_\pi$ and  $\gamma_\pi$ satisfy $\log\{ {\pi(x)}/{(1-\pi(x))} \} = \lambda_\pi^\mathsf{T}H(x) + \gamma_\pi^\mathsf{T}G(x)$.
        \item $\log\{ {\pi(x)}/{(1-\pi(x))} \} \in \textnormal{Span}\{H(x), G(x)\}$ and $\tau(x) \in \textnormal{Span}\{H(x)\}$.
    \end{enumerate}
\end{theorem}

Even though Conditions (a)-(c) are quite strong, the consistency in Theorem \ref{thm:consistency_ebal_HG} requires only one of the Conditions (a)-(c). 
Conditions (a) and (b) are similar to the requirement for the double robustness property established in the previous entropy balancing works \citep{zhao2016entropy,dong2020integrative,josey2020calibration}. In particular, Condition (a) assumes that the means of the potential outcomes follow certain forms, and Condition (b) puts some restrictions on treatment assignment and density ratio between the two populations. Condition (c) identifies another situation in which consistent estimation can be achieved. Although this condition contains both assumptions of treatment assignment and potential outcomes regression, it is not stronger than either Condition (a) or (b). The first part of Condition (c) is milder than Condition (b) as it does not restrict the density ratio between the two populations. The second part of Condition (c) is milder than Condition (a) as it only requires the  means of the effect modifiers from the target population contained in $H$. Intuitively, since the difference in ATEs between the source and the target populations arises from their differences in effect modifier distributions, calibrating the estimate can only be successful when there is sufficient information about the effect modifiers from the target population. In this sense, the second part of Condition (c) is a minimal requirement for consistency.

As \eqref{eq:ebal_only_H} is a special case of \eqref{eq:ebal_HG} with $G$ being a null set, the theorem also applies to the estimates from \cite{josey2020calibration}. In this case, Conditions (b) and (c) require   $\log\{ {\pi(x)}/{(1-\pi(x))}$ to be linear in $\{h_k\}$. These conditions become more stringent as $H$ may be conveniently chosen and may not contain all the important confounders. On the contrary, the additional balancing constraints in \eqref{eq:ebal_HG} create the possibility of adjusting for confounding on an additional set of covariates $G$ that can be carefully chosen.

\begin{theorem}[Asymptotic variance]
    \label{thm:var_EEB}
	Suppose $\hat{w}$ is the solution of \eqref{eq:ebal_HG}. Assume either Conditions (b) or (c) in Theorem \ref{thm:consistency_ebal_HG} holds. Then the asymptotic variance of $\hat{\tau}_{\hat{w}}$ with $\hat{w}$ given by \eqref{eq:ebal_HG} is
    \begin{equation}
        \label{eq:asym_var}
        \begin{aligned}
			&
			\frac{1}{\rho^2} 
			\E \left[ S\tilde{r}(X)^2 \left\{ 
				\frac{\sigma^2_1(X)}{\pi(X)}  +
				\frac{\sigma^2_0(X)}{1-\pi(X)} 
			\right\} \right]
			+
			\frac{1}{(1-\rho)^2} \E \Big[ (1-S)\left\{ \Pi_H(\tau(X)) - \tau^* \right\}^2 \Big]
			\\&+
			\frac{1}{\rho^2} \E \Bigg[ \frac{S \tilde{r}(X)^2}{\pi(X)} \left\{ \mu_1(X) - \Pi_H(\mu_1(X)) - \Pi_{G^\perp}(m(X)) \right\}^2 
			\\&\hspace{3em}+
			\frac{S \tilde{r}(X)^2}{1-\pi(X)} \left\{ \mu_0(X) - \Pi_H(\mu_0(X)) - \Pi_{G^\perp}(m(X)) \right\}^2 \Bigg]
		.\end{aligned}
    \end{equation}
	Here $\rho = \E[\rho(X)]$, $\tilde{r}(x)$ is defined in \eqref{eq:tilde_r}, $\sigma^2_a(X) = \var(Y(a) \mid X, S=1)$ and $m(X) = {(\mu_1(X) + \mu_0(X))}/{2}$. $\Pi_H(\cdot)$ denotes projection on $\textnormal{Span}\{H(X)\}$ w.r.t. covariate distribution $p_s(x)\tilde{r}(x)$, where $p_s(x)$ is the covariate distribution in the source population. Specifically, 
	\begin{equation*}
	\Pi_H\{\tau(X)\} = \E\{\tilde{r}(X) \tau(X) H(X)^\mathsf{T} \mid S = 1\} \E \{\tilde{r}(X) H(X)H(X)^\mathsf{T} \mid S = 1 \}^{-1} H(X)
	.\end{equation*}
	$\Pi_H(\mu_1(X))$ and $\Pi_H(\mu_0(X))$ are defined in a similar manner. $\Pi_{G^\perp}(\cdot)$ denotes projection on $\textnormal{Span}\{G^\perp(X)\}$, where $G^\perp(X) = G(X) - \Pi_H(G(X))$.
\end{theorem}

The first term of the asymptotic variance \eqref{eq:asym_var} is equal to 
$$\frac{1}{(1-\rho)^2} \E\left[ \frac{(1-\rho(X))^2}{\rho(X)} \left\{ \frac{\sigma^2_1(X)}{\pi(X)} + \frac{\sigma^2_0(X)}{1-\pi(X)} \right\} \right]$$ 
when Condition (b) holds. If Condition (c) holds, obviously $\Pi_H(\tau(X)) = \tau(X)$. Therefore, when Conditions (b) and (c) both hold, the sum of first two terms of \eqref{eq:asym_var} becomes
\begin{align*}
	&\frac{1}{(1-\rho)^2} \E\left[ \frac{(1-\rho(X))^2}{\rho(X)} \left\{ \frac{\sigma^2_1(X)}{\pi(X)} + \frac{\sigma^2_0(X)}{1-\pi(X)} \right\} \right] 
 + 
	\frac{1}{(1-\rho)^2} \E\left[ (1-\rho(X)) (\tau(X) - \tau^*)^2 \right]
.\end{align*}
This is exactly the semiparametric efficiency bound for estimating $\tau^*$ if individual-level target sample data is available. Note that the third term of \eqref{eq:asym_var} is always non-negative, so when Conditions (b) and (c) hold, this term quantifies how much the asymptotic variance exceeds the efficiency bound. If Condition (a) also holds, it is easy to check that the third term vanishes and thus the efficiency bound is achieved by the proposed method.

The asymptotic variance also characterizes the efficiency gain from the additional balancing constraints between the treated and control groups on $G$. When Condition (c) holds, $\Pi_{G^\perp}\{\tau(X)\} = 0$. Then by $m(X) = \mu_1(X) - {\tau(X)}/{2}$, we have $\Pi_{G^\perp}\{m(X)\} = \Pi_{G^\perp}\{\mu_1(X)\}$. Hence, 
\begin{align*}
		\setstretch{1}
\big[\mu_1(X) - \Pi_H\{\mu_1(X)\} - \Pi_{G^\perp}\{m(X)\} \big]^2 
&= 
\big[ \mu_1(X) - \Pi_H\{\mu_1(X)\} - \Pi_{G^\perp}\{\mu_1(X)\} \big]^2 
\\&= 
\big[ \mu_1(X) - \Pi_{H + G}\{\mu_1(X)\} \big]^2
,\end{align*}
where $\Pi_{H + G}(\cdot)$ the projection on $\textnormal{Span}\{H(X), G(X)\}$. Similarly, $m(X) = \mu_0(X) + {\tau(X)}/{2}$ implies $\big[ \mu_0(X) - \Pi_H\{\mu_0(X)\} - \Pi_{G^\perp}\{m(X)\} \big]^2 =\big[ \mu_0(X) - \Pi_{H + G}\{\mu_0(X)\} \big ]^2$. 
Since $\big[ \mu_a(X) - \Pi_{H + G}\{\mu_a(X)\} \big]^2 \le \big[ \mu_a(X) - \Pi_{H}\{\mu_a(X)\} \big]^2,$ we conclude that \eqref{eq:ebal_HG} may result in a more efficient estimation than \eqref{eq:ebal_only_H} even when some (or all) elements in $G$ are not related to the underlying true propensity score. 

Such a result is not surprising --- it has been well noted in the causal inference literature that inclusion of covariates that affect the outcomes but are not necessarily related to treatment assignment can lead to efficiency gains \citep{brookhart2006variable,shortreed2017outcome}. This result also suggests that the proposed method can also improve the precision of generalizing effect estimation from an RCT to a target population, especially when the outcomes are largely affected by some covariates. Although by randomization these covariates follow the same distribution across the treated and control groups, under finite samples covariate imbalance often occurs due to sampling error \citep{li2020rerandomization}. The proposed method provides a strategy to tackle such imbalance without an explicit regression adjustment. Lastly, even though our theoretical result quantifies the asymptotic efficiency gain from balancing additional terms, it can not be directly used to make inference about the target population ATE under our setting. To see why, observe that the expectations in the asymptotic variance formula \eqref{eq:asym_var} are based on the joint distribution of $(X,S)$ which involve both the source and target populations. In particular, the term $\mathbb{E} \Big[ (1-S)\left\{ \Pi_H(\tau(X)) - \tau^* \right\}^2 \Big]$ needs individual level data from the target population for estimation because of $(1-S)$ inside the expectation. Since we only have summary level covariate information for the target population, the variance of $\hat{\tau}_{w}$ ($\text{Var}(\hat{\tau}_{w})$) cannot be numerically evaluated without further assumptions. In the Web Appendix, we propose an upper bound for $\text{Var}(\hat{\tau}_{w})$ under extra conditions, and we plan to fully investigate such strategy and alternative strategies in the future. In general, constructing valid confidence interval of $\tau^{*}$ is still an open problem, and it could be an interesting future direction.

\section{Simulation studies}
\label{sec:simu}
In this section, we conduct simulation studies to evaluate the performance of the proposed method in finite sample settings. In our simulation setup, we generate covariates $X = (X_1, \ldots, X_5)$ from a uniform distribution on $[-2, 2]^5$. The participation probability is set as $\text{logit} \{\rho(x) \} =  0.4 x_1 + 0.3 x_2 - 0.2 x_4$ so that $(X_1, X_2, X_4)$ have different distributions between the source and the target populations. We set $H(x) = (1, x_1, x_2, x_3)$ so that  the only available information from the target sample is the sample means of $X_1, X_2$ and $X_3$. We consider balancing on the first moments of all covariates, so $G(x) = (x_4, x_5)$.

To test the performance under various scenarios, we consider three propensity score models for treatment assignment:
\begin{enumerate}[label=(P\arabic*), leftmargin=6ex]
	\item $\text{logit} \{\pi(x)\} =0.7 x_2 + 0.5 x_3$ \hfill{(linear, only related to $\{h_k\}$)}
	\item $\text{logit} \{\pi(x)\} = 0.35 x_2 + 0.25 x_3 + 0.2 x_4 - 0.7 x_5$ \hfill{(linear, related to $\{h_k\}$ and $\{g_k\}$)}
	\item $\text{logit} \{\pi(x)\} =0.35 x_2 - 0.4 \max(x_3, x_4) - 0.7 x_5$ (nonlinear, related to $\{h_k\}$ and $\{g_k\}$).
\end{enumerate}
In the first scenario, the logit of the propensity score is linear in $H$, so covariate balancing on $H$ suffices to account for confounding. In the last two scenarios, the propensity score is also related to $G$. Furthermore, the third setting contains a nonlinear term.

The observed outcomes in the source sample $Y_i = m(X_i) + (A_i - 0.5) \tau(X_i) + \varepsilon_i$ with $\varepsilon_i \distas{i.i.d.} N(0, 1)$. Two settings for the CATE function $\tau(x)$ are considered.
\begin{enumerate}[label=(T\arabic*), leftmargin=6ex]
	\item $\tau(x) = x_1 - 0.6 x_2 - 0.4 x_3$,
	\item $\tau(x) = x_1 - 0.5 \exp(x_2 - 0.5 x_3)$.
\end{enumerate}
Under (T1), $\tau(x)$ is linear in $H(x)$ hence satisfies Condition (c). But this condition does not hold under (T2). Finally we consider the following choices for $m(x)$:
\begin{enumerate}[label=(M\arabic*), leftmargin=6ex]
	\item $m(x) = 0.5 x_1 + 0.3 x_2 + 0.3 x_3 - 0.4 x_4 - 0.5 x_5$,
	\item $m(x) = 0.5 x_1 + 0.3 x_2^2 + 0.2 \exp(x_3 - x_4 - 1) - 0.5 x_5$.
\end{enumerate}
Under the setting (T1)+(M1), Condition (a) in Theorem \ref{thm:consistency_ebal_HG} is satisfied. In contrast, under the other scenarios, this condition no longer holds.

For comparison, we consider other methods for constructing weights, including both modeling and balancing approaches. First, we consider the conventional inverse probability weight (IPW) based on modeling the propensity score. We use the source sample to fit a logistic regression for $A$ against all the covariates, including $H(x)$ and $G(x)$. Then the weights are computed as $1/\hat{\pi}(X_i)$ for the treated units and $1/\{1 - \hat{\pi}(X_i)\}$ for the control units. This method is labeled as \texttt{IPW}. Note that the IPW approach does not incorporate the sample means of the target sample to adjust for covariate shift. As discussed in Section \ref{sec:method_review}, the covariate shift adjustment approach of \citet{dong2020integrative} is equivalent to modeling $\rho(x) / \{1-\rho(x)\}$ with exponential tilting, so we can combine it with the IPW method. Specifically, we compute $\hat{q}_i$ by solving \eqref{eq:ebal_dong}, and then set the weights as $\hat{q}_i / \hat{\pi}(X_i)$ for the treated units and $\hat{q}_i / \{1 - \hat{\pi}(X_i)\}$ for the control units. This method is labeled as \texttt{IPW+ET}, where ``ET'' stands for exponential tilting. The last comparator method, labeled as \texttt{EBAL}, uses the weights proposed by \citet{josey2020calibration}, which are given by \eqref{eq:ebal_only_H}. Each set of weights is normalized so that the sums of the weights within the treated and control groups are both equal to $n_s$. ATEs are estimated from \eqref{eq:tau_hat_w}. The total sample size is set as $n = 800$. The size of source sample $n_s$ varies from $340$ to $430$ in our simulations due to variation in the participation probability.

\begin{figure}[!tbp]
    \centering
    \includegraphics[width=.85\linewidth]{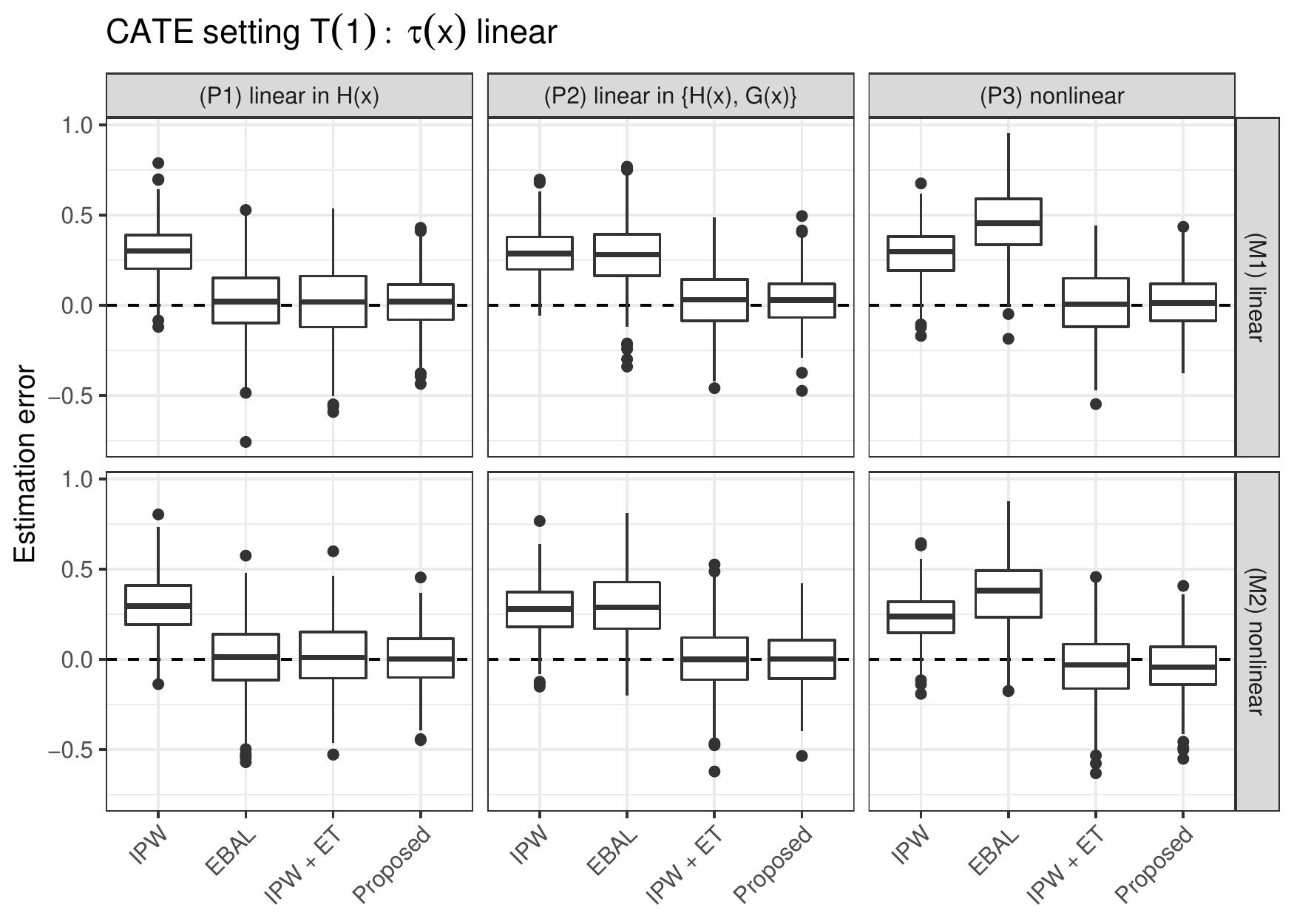}
    \includegraphics[width=.85\linewidth]{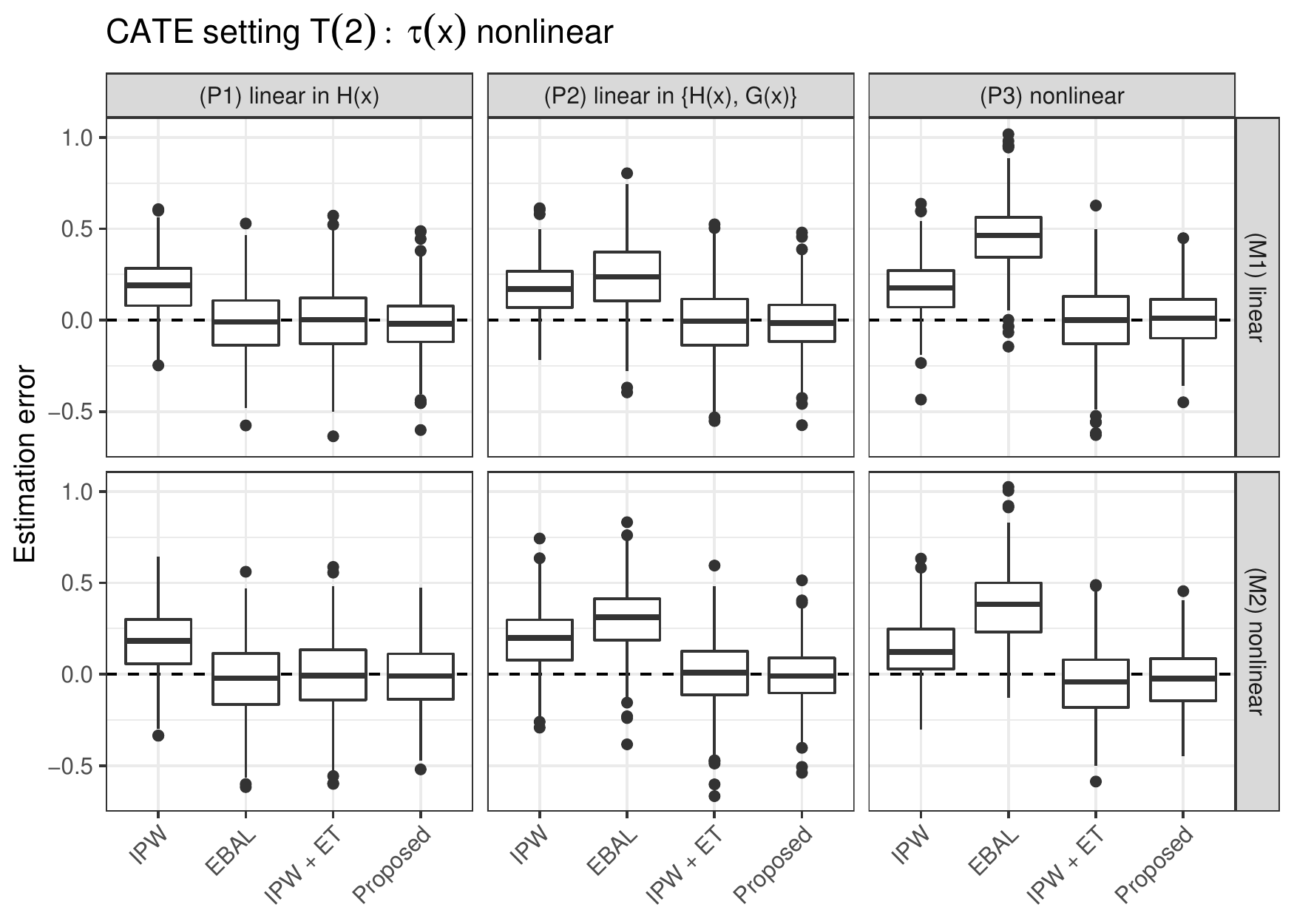}
	\caption{Boxplots of estimation errors under various simulation scenarios.}
    \label{fig:simu_err}
\end{figure}

The performance is measured in terms of estimation error with respect to $\tau^*$. Figure \ref{fig:simu_err} plots the estimation errors over 400 independent runs in boxplots. Overall, the performance pattern of different methods under linear CATE is similar to that under nonlinear CATE. The conventional IPW gives biased results under all scenarios because it does not account for covariate shift. Although $X_4$ has different distribution across the source population and the target population, only the covariate shift on the effect modifiers, i.e., $X_1$ and $X_2$, needs to be adjusted. The entropy balancing weights given by \eqref{eq:ebal_only_H} can serve this end and adjust for confounding simultaneously when all the confounders are contained in $H$ for Setting (P1). However, if $H$ misses some confounders in Settings (P2) or (P3), \texttt{EBAL} no longer gives consistent estimates. In contrast, our extended entropy balancing strategy is able to retain consistency under a wider range of situations. Moreover, when the consistency condition for \eqref{eq:ebal_only_H} holds, the proposed method can achieve higher efficiency. The efficiency improvement comes from balancing on covariates that are related to the potential outcomes even though they are not in the propensity score ($X_4$ and $X_5$ in this case).

Although \texttt{IPW+ET} can also achieve consistent estimation in many scenarios like the proposed method, the result is not as efficient as the proposed method. In a number of simulation runs, this method produces rather unstable estimates, even when the propensity score models are perfectly specified. Such instability is because inverting the probability estimates could potentially inflate the estimation error when the estimated probability is close to 0 or 1. In contrast, the balancing-based weighting approaches generally lead to more stable estimation. In the Web Appendix, we provide additional simulations when none of Conditions (a)-(c) hold. We see that while all the competing methods are no longer consistent, the proposed one usually gives the smallest bias.
\section{Application: TTEC and mortality in sepsis}
\label{sec:data}

In this section, we illustrate the proposed method for evaluating the treatment effect of transthoracic echocardiography (TTEC) for intensive care unit (ICU) patients with sepsis. We use the same dataset in \citet{feng2018transthoracic}, which is derived from the MIMIC-III database \citep{johnson2016mimic}. This is an observational dataset consisting of 6361 ICU patients. Among the patients, 51.3\% had TTEC performed during or in the period less than 24 hours before their ICU admission. We use 28-day survival as the outcome. The dataset contains 17 baseline variables: age (range 18 - 91), gender, weight; severity at admission, which is measured by the simplified acute physiology score (SAPS), the sequential organ failure assessment (SOFA) score, the Elixhauser comorbidity score, comorbidity indicators, including congestive heart failure, atrial fibrillation, respiratory failure, malignant tumor, vital signs, including mean arterial pressure, heart rate and temperature; laboratory results, including platelet count, partial pressure of oxygen, lactate, and blood urea nitrogen.
The distributions of lab results are right-skewed, so we apply log transformations on these variables. All the continuous variables are then standardized for further analysis. We impute any missing values using MissForest \citep{stekhoven2012missforest}, which is a flexible non-parametric missing value imputation approach.

We assume only the average values of the demographic covariates (age, gender, weight) and comorbidity indicators (congestive heart failure, atrial fibrillation, respiratory failure, malignant tumor) are available from the target population.

In order to evaluate the generalization performance under a wide range of settings, we use the following sampling design to induce different levels of covariate shift and confounding, but preserve the covariate-outcome relationship in the real data. First, we sample 40\% of the data to construct the source sample, and the remaining data is evenly, randomly broken into a target sample and a test sample. In this way, the test sample follows the same distribution as the target sample. The probability of being selected as a source sample is proportional to
\begin{equation*}
	\Psi\{ \kappa_S (-0.3 \times \texttt{age} + 0.3 \times \texttt{cmb}_1 + 0.4 \times \texttt{cmb}_2 + 0.3 \times \texttt{cmb}_3 + 0.4 \times \texttt{cmb}_4 - 0.5) \}
,\end{equation*}
where $\Psi(z) = 0.8\Phi(z) + 0.1$ and $\Phi(z)$ is the standard normal cumulative probability function. Here \texttt{cmb}$_i$'s represent the comorbidity indicators and $\kappa_S$ is a parameter to induce different levels of covariate shift. We set $\kappa_S=1$ for small covariate shift and $\kappa_S=5$ for large covariate shift. Under this sampling design, the target population is older and more likely to have comorbidities than the source population.
Among the 40\% of the sampled data, we randomly select 50\% of the TTEC patients and 50\% of the non-TTEC patients to form the source sample. The treated units are selected with probability proportional to $g(X_i)$ while the control units are with probability proportional to $0.5 - g(X_i)$, where $g$ is set as 
\begin{equation*}
	g(x) = \Psi( \kappa_A (0.3 \times \texttt{SAPS} + 0.4 \times \texttt{SOFA} - 0.5\times \texttt{Elixhauser}))
.\end{equation*}
We consider two choices of $\kappa_A$: (a) $\kappa_A = 0$, so all the patients in this step are sampled with equal probability; (b) $\kappa_A=1$, which induces additional confounding determined by a linear combination of the severity scores.

\begin{table*}[!tbp]
    \small
    \centering
    \caption{Data analysis results (biases and RMSEs are multiplied by 100).}
    \label{tab:data}
	\vspace{2pt}
    \begin{tabular}{
        l@{\hspace{20pt}}
        S[detect-weight]@{\hspace{5pt}}S[detect-weight]c@{\hspace{14pt}}
        S[detect-weight]@{\hspace{5pt}}S[detect-weight]c@{\hspace{14pt}}
        S[detect-weight]@{\hspace{5pt}}S[detect-weight]c@{\hspace{14pt}}
        S[detect-weight]@{\hspace{5pt}}S[detect-weight]
    }
    \toprule
    \multirow{2}{*}{Method} & 
    \multicolumn{2}{c}{Setting 1} & &
    \multicolumn{2}{c}{Setting 2} & &
    \multicolumn{2}{c}{Setting 3} & &
    \multicolumn{2}{c}{Setting 4} 
    \vspace{-1pt} \\ \cmidrule(lr){2-3} \cmidrule(lr){5-6} \cmidrule(lr){8-9} \cmidrule(lr){11-12} \vspace{-13pt}\\ \vspace{-1pt}
    & \multicolumn{1}{l}{Bias} & \multicolumn{1}{c}{RMSE} &
    & \multicolumn{1}{l}{Bias} & \multicolumn{1}{c}{RMSE} &
    & \multicolumn{1}{l}{Bias} & \multicolumn{1}{c}{RMSE} &
    & \multicolumn{1}{l}{Bias} & \multicolumn{1}{c}{RMSE}
	\\ \midrule 
	  IPW & 1.33 & 3.35 && 3.36 & 4.99 && 3.05 & 4.32 && 5.34 & 6.46 \\
      EBAL & -2.22 & 3.87 && -6.92 & 7.61 && -1.85 & 4.81 && -6.96 & 8.22 \\
      IPW + ET & 0.30 & 3.18 && 2.09 & 4.19 && \bfseries 0.37 & 4.86 && 1.89 & 5.27 \\
      proposed & \bfseries0.13 & \bfseries3.09 && \bfseries -0.02 & \bfseries 3.30 && 0.51 & \bfseries 4.30 && \bfseries 0.21 & \bfseries 4.24 \\
  \bottomrule
    \end{tabular}
	\begin{tablenotes}
	\item 
	Setting 1 ($\kappa_S=1, ~ \kappa_A=0$): small covariate shift, no extra confounding;\\
	Setting 2 ($\kappa_S=1, ~ \kappa_A=1$): small covariate shift, extra confounding;\\
	Setting 3 ($\kappa_S=5, ~ \kappa_A=0$): large covariate shift, no extra confounding;\\
	Setting 4 ($\kappa_S=5, ~ \kappa_A=1$): large covariate shift, extra confounding.
	\end{tablenotes}
\end{table*}

So in total we have considered 4 settings; under each one, we run 800 replications of the sampling procedure. For each replication, we apply the proposed method and the comparator methods in Section \ref{sec:simu} to estimate the ATE for the target population. To obtain a benchmark for these results, we compute an oracle estimate of the target population ATE by constructing the entropy balancing weights based ATE estimator using the full information in the test data, which contains the observed treatment, the observed outcome and the individual-level covariates information. Then the estimation errors are recorded as the difference between the generalization estimates and the oracle estimate. Table \ref{tab:data} summarizes the mean value of the estimation errors as bias, and their root mean square as ``rmse''. As we can see, the proposed method usually achieves the smallest bias and rmse among all the weighting methods, especially when there is a substantial difference between the source and target populations. Note that such an information from the test data is not needed to apply our proposed methods (as well as comparator methods), it is included here merely for results evaluation.

\section{Discussion}
\label{sec:discussion}

In this paper, we have proposed a covariate balancing weighting approach for estimating the ATE for a given target population under the situation where only summary-level data from a target sample is available. The proposed method is motivated by the recently developed entropy balancing methods \citep{hainmueller2012entropy,dong2020integrative,josey2020calibration}, but is extended with additional treatment-control balancing terms over a broader set of covariate functions. The consistency conditions and asymptotic variance of the corresponding weighting estimator are characterized. 

Throughout this paper we have focused on seeking covariate balancing on two fixed sets of functions $H$ and $G$. In practice, while the choice of $H$ is determined by what information is available from the target sample, the specification of $G$ can be a subtle issue. If $G$ contains too few functions, it may not be rich enough to account for confounding; on the other hand, if $G$ contains too many functions, the linear constraints in \eqref{eq:ebal_HG} may not admit a feasible solution. There are a few extensions that we can consider to this end. First, we may allow the dimension of $G$ to grow with the source sample size, for example, using the method of sieves, similar to \citet{chan2016globally}. Also, to ensure the feasibility of the balancing constraints, we can seek approximate balance instead of exact balance. \citet{wang2020minimal} showed that by allowing small imbalance, one could incorporate more covariate functions into the balancing constraints and lead to possibly preferable results. Alternatively, we can utilize the kernel balancing idea in \citet{wong2018kernel} and minimize the treatment-control imbalance over a non-parametric function class. These are all interesting future directions to explore.

 \section*{Acknowledgments}
The authors thank the referees and the Associate Editor for their insightful comments and suggestions that greatly improved the paper. Research reported in this work was partially funded through a Patient-Centered Outcomes Research Institute (PCORI) Award (ME-2018C2-13180). The views in this work are solely the responsibility of the authors and do not necessarily represent the views of the Patient-Centered
Outcomes Research Institute (PCORI), its Board of Governors or Methodology Committee.\\

\section*{Data Availability Statement}
The data that support the findings of this study were derived from the following resources available in the public domain: MIMIC-III Clinical Database Version 1.4 (\url{https://physionet.org/content/mimiciii/1.4/})

\begingroup
\bibliographystyle{biom} 
\bibliography{ref_generalizability}

\begin{thebibliography}{}

\bibitem[\protect\citeauthoryear{Bennett, Vielma, and Zubizarreta}{Bennett
  et~al.}{2020}]{bennett2020building}
Bennett, M., Vielma, J.~P., and Zubizarreta, J.~R. (2020).
\newblock Building representative matched samples with multi-valued treatments
  in large observational studies.
\newblock {\em Journal of Computational and Graphical Statistics} {\bf 29,}
  744--757.

\bibitem[\protect\citeauthoryear{Brookhart, Schneeweiss, Rothman, Glynn, Avorn,
  and St{\"u}rmer}{Brookhart et~al.}{2006}]{brookhart2006variable}
Brookhart, M.~A., Schneeweiss, S., Rothman, K.~J., Glynn, R.~J., Avorn, J., and
  St{\"u}rmer, T. (2006).
\newblock Variable selection for propensity score models.
\newblock {\em American Journal of Epidemiology} {\bf 163,} 1149--1156.

\bibitem[\protect\citeauthoryear{Buchanan, Hudgens, Cole, Mollan, Sax, Daar,
  Adimora, Eron, and Mugavero}{Buchanan
  et~al.}{2018}]{buchanan2018generalizing}
Buchanan, A.~L., Hudgens, M.~G., Cole, S.~R., Mollan, K.~R., Sax, P.~E., Daar,
  E.~S., Adimora, A.~A., Eron, J.~J., and Mugavero, M.~J. (2018).
\newblock Generalizing evidence from randomized trials using inverse
  probability of sampling weights.
\newblock {\em Journal of the Royal Statistical Society: Series A (Statistics
  in Society)} {\bf 181,} 1193--1209.

\bibitem[\protect\citeauthoryear{Chan, Yam, and Zhang}{Chan
  et~al.}{2016}]{chan2016globally}
Chan, K. C.~G., Yam, S. C.~P., and Zhang, Z. (2016).
\newblock Globally efficient non-parametric inference of average treatment
  effects by empirical balancing calibration weighting.
\newblock {\em Journal of the Royal Statistical Society: Series B (Statistical
  Methodology)} {\bf 78,} 673--700.

\bibitem[\protect\citeauthoryear{Chattopadhyay, Hase, and
  Zubizarreta}{Chattopadhyay et~al.}{2020}]{chattopadhyay2020balancing}
Chattopadhyay, A., Hase, C.~H., and Zubizarreta, J.~R. (2020).
\newblock Balancing vs modeling approaches to weighting in practice.
\newblock {\em Statistics in Medicine} {\bf 39,} 3227--3254.

\bibitem[\protect\citeauthoryear{Cole and Stuart}{Cole and
  Stuart}{2010}]{Cole2010}
Cole, S.~R. and Stuart, E.~A. (2010).
\newblock {Generalizing Evidence From Randomized Clinical Trials to Target
  Populations: The ACTG 320 Trial}.
\newblock {\em American Journal of Epidemiology} {\bf 172,} 107--115.

\bibitem[\protect\citeauthoryear{Colnet, Mayer, Chen, Dieng, Li, Varoquaux,
  Vert, Josse, and Yang}{Colnet et~al.}{2020}]{colnet2020causal}
Colnet, B., Mayer, I., Chen, G., Dieng, A., Li, R., Varoquaux, G., Vert, J.-P.,
  Josse, J., and Yang, S. (2020).
\newblock Causal inference methods for combining randomized trials and
  observational studies: a review.
\newblock {\em arXiv preprint arXiv:2011.08047} .

\bibitem[\protect\citeauthoryear{Dahabreh and Hern{\'a}n}{Dahabreh and
  Hern{\'a}n}{2019}]{dahabreh2019extending}
Dahabreh, I.~J. and Hern{\'a}n, M.~A. (2019).
\newblock Extending inferences from a randomized trial to a target population.
\newblock {\em European journal of epidemiology} {\bf 34,} 719--722.

\bibitem[\protect\citeauthoryear{Dahabreh, Robertson, Steingrimsson, Stuart,
  and Hernan}{Dahabreh et~al.}{2020}]{dahabreh2020extending}
Dahabreh, I.~J., Robertson, S.~E., Steingrimsson, J.~A., Stuart, E.~A., and
  Hernan, M.~A. (2020).
\newblock Extending inferences from a randomized trial to a new target
  population.
\newblock {\em Statistics in Medicine} {\bf 39,} 1999--2014.

\bibitem[\protect\citeauthoryear{Degtiar and Rose}{Degtiar and
  Rose}{2021}]{degtiar2021review}
Degtiar, I. and Rose, S. (2021).
\newblock A review of generalizability and transportability.
\newblock {\em arXiv preprint arXiv:2102.11904} .

\bibitem[\protect\citeauthoryear{Dong, Yang, Wang, Zeng, and Cai}{Dong
  et~al.}{2020}]{dong2020integrative}
Dong, L., Yang, S., Wang, X., Zeng, D., and Cai, J. (2020).
\newblock Integrative analysis of randomized clinical trials with real world
  evidence studies.
\newblock {\em arXiv preprint arXiv:2003.01242} .

\bibitem[\protect\citeauthoryear{Feng, McSparron, Kien, Stone, Roberts,
  Schwartzstein, Vieillard-Baron, and Celi}{Feng
  et~al.}{2018}]{feng2018transthoracic}
Feng, M., McSparron, J.~I., Kien, D.~T., Stone, D.~J., Roberts, D.~H.,
  Schwartzstein, R.~M., Vieillard-Baron, A., and Celi, L.~A. (2018).
\newblock Transthoracic echocardiography and mortality in sepsis: analysis of
  the mimic-iii database.
\newblock {\em Intensive Care Medicine} {\bf 44,} 884--892.

\bibitem[\protect\citeauthoryear{Hahn}{Hahn}{1998}]{hahn1998role}
Hahn, J. (1998).
\newblock On the role of the propensity score in efficient semiparametric
  estimation of average treatment effects.
\newblock {\em Econometrica} {\bf 66,} 315--331.

\bibitem[\protect\citeauthoryear{Hainmueller}{Hainmueller}{2012}]{hainmueller2012entropy}
Hainmueller, J. (2012).
\newblock Entropy balancing for causal effects: A multivariate reweighting
  method to produce balanced samples in observational studies.
\newblock {\em Political Analysis} {\bf 20,} 25--46.

\bibitem[\protect\citeauthoryear{Hartman, Grieve, Ramsahai, and Sekhon}{Hartman
  et~al.}{2015}]{hartman2015sample}
Hartman, E., Grieve, R., Ramsahai, R., and Sekhon, J.~S. (2015).
\newblock From sample average treatment effect to population average treatment
  effect on the treated: combining experimental with observational studies to
  estimate population treatment effects.
\newblock {\em Journal of the Royal Statistical Society. Series A (Statistics
  in Society)} {\bf 178,} 757--778.

\bibitem[\protect\citeauthoryear{Hong, Webster-Clark, Jonsson~Funk,
  St{\"u}rmer, Dempster, Cole, Herr, and LoCasale}{Hong
  et~al.}{2019}]{hong2019comparison}
Hong, J.-L., Webster-Clark, M., Jonsson~Funk, M., St{\"u}rmer, T., Dempster,
  S.~E., Cole, S.~R., Herr, I., and LoCasale, R. (2019).
\newblock Comparison of methods to generalize randomized clinical trial results
  without individual-level data for the target population.
\newblock {\em American Journal of Epidemiology} {\bf 188,} 426--437.

\bibitem[\protect\citeauthoryear{Imai and Ratkovic}{Imai and
  Ratkovic}{2014}]{imai2014covariate}
Imai, K. and Ratkovic, M. (2014).
\newblock Covariate balancing propensity score.
\newblock {\em Journal of the Royal Statistical Society: Series B (Statistical
  Methodology)} {\bf 76,} 243--263.

\bibitem[\protect\citeauthoryear{Johnson, Pollard, Shen, Li-Wei, Feng,
  Ghassemi, Moody, Szolovits, Celi, and Mark}{Johnson
  et~al.}{2016}]{johnson2016mimic}
Johnson, A.~E., Pollard, T.~J., Shen, L., Li-Wei, H.~L., Feng, M., Ghassemi,
  M., Moody, B., Szolovits, P., Celi, L.~A., and Mark, R.~G. (2016).
\newblock Mimic-iii, a freely accessible critical care database.
\newblock {\em Scientific Data} {\bf 3,} 1--9.

\bibitem[\protect\citeauthoryear{Josey, Yang, Ghosh, and Raghavan}{Josey
  et~al.}{2020}]{josey2020calibration}
Josey, K.~P., Yang, F., Ghosh, D., and Raghavan, S. (2020).
\newblock A calibration approach to transportability with observational data.
\newblock {\em arXiv preprint arXiv:2008.06615} .

\bibitem[\protect\citeauthoryear{Kang, Schafer, et~al\mbox{.}}{Kang
  et~al.}{2007}]{kang2007demystifying}
Kang, J.~D., Schafer, J.~L., et~al. (2007).
\newblock Demystifying double robustness: A comparison of alternative
  strategies for estimating a population mean from incomplete data.
\newblock {\em Statistical Science} {\bf 22,} 523--539.

\bibitem[\protect\citeauthoryear{Li and Ding}{Li and
  Ding}{2020}]{li2020rerandomization}
Li, X. and Ding, P. (2020).
\newblock Rerandomization and regression adjustment.
\newblock {\em Journal of the Royal Statistical Society: Series B (Statistical
  Methodology)} {\bf 82,} 241--268.

\bibitem[\protect\citeauthoryear{Lu, Ben-Michael, Feller, and Miratrix}{Lu
  et~al.}{2021}]{lu2021you}
Lu, B., Ben-Michael, E., Feller, A., and Miratrix, L. (2021).
\newblock Is it who you are or where you are? accounting for compositional
  differences in cross-site treatment variation.
\newblock {\em arXiv preprint arXiv:2103.14765} .

\bibitem[\protect\citeauthoryear{Rosenbaum and Rubin}{Rosenbaum and
  Rubin}{1983}]{rosenbaum1983central}
Rosenbaum, P.~R. and Rubin, D.~B. (1983).
\newblock The central role of the propensity score in observational studies for
  causal effects.
\newblock {\em Biometrika} {\bf 70,} 41--55.

\bibitem[\protect\citeauthoryear{Rothwell}{Rothwell}{2005}]{Rothwell2005}
Rothwell, P.~M. (2005).
\newblock External validity of randomised controlled trials: “to whom do the
  results of this trial apply?”.
\newblock {\em The Lancet} {\bf 365,} 82 -- 93.

\bibitem[\protect\citeauthoryear{Rubin}{Rubin}{1974}]{rubin1974estimating}
Rubin, D.~B. (1974).
\newblock Estimating causal effects of treatments in randomized and
  nonrandomized studies.
\newblock {\em Journal of educational Psychology} {\bf 66,} 688.

\bibitem[\protect\citeauthoryear{Rudolph and van~der Laan}{Rudolph and van~der
  Laan}{2017}]{Rudolph2017}
Rudolph, K.~E. and van~der Laan, M.~J. (2017).
\newblock Robust estimation of encouragement-design intervention effects
  transported across sites.
\newblock {\em Journal of the Royal Statistical Society: Series B (Statistical
  Methodology)} {\bf 79,} 1509--1525.

\bibitem[\protect\citeauthoryear{Shortreed and Ertefaie}{Shortreed and
  Ertefaie}{2017}]{shortreed2017outcome}
Shortreed, S.~M. and Ertefaie, A. (2017).
\newblock Outcome-adaptive lasso: Variable selection for causal inference.
\newblock {\em Biometrics} {\bf 73,} 1111--1122.

\bibitem[\protect\citeauthoryear{Signorovitch, Wu, Andrew, Gerrits, Kantor,
  Bao, Gupta, and Mulani}{Signorovitch
  et~al.}{2010}]{signorovitch2010comparative}
Signorovitch, J.~E., Wu, E.~Q., Andrew, P.~Y., Gerrits, C.~M., Kantor, E., Bao,
  Y., Gupta, S.~R., and Mulani, P.~M. (2010).
\newblock Comparative effectiveness without head-to-head trials.
\newblock {\em Pharmacoeconomics} {\bf 28,} 935--945.

\bibitem[\protect\citeauthoryear{Silber, Rosenbaum, Ross, Ludwig, Wang, Niknam,
  Mukherjee, Saynisch, Even-Shoshan, Kelz, et~al\mbox{.}}{Silber
  et~al.}{2014}]{silber2014template}
Silber, J.~H., Rosenbaum, P.~R., Ross, R.~N., Ludwig, J.~M., Wang, W., Niknam,
  B.~A., Mukherjee, N., Saynisch, P.~A., Even-Shoshan, O., Kelz, R.~R., et~al.
  (2014).
\newblock Template matching for auditing hospital cost and quality.
\newblock {\em Health services research} {\bf 49,} 1446--1474.

\bibitem[\protect\citeauthoryear{Stekhoven and B{\"u}hlmann}{Stekhoven and
  B{\"u}hlmann}{2012}]{stekhoven2012missforest}
Stekhoven, D.~J. and B{\"u}hlmann, P. (2012).
\newblock Missforest—non-parametric missing value imputation for mixed-type
  data.
\newblock {\em Bioinformatics} {\bf 28,} 112--118.

\bibitem[\protect\citeauthoryear{Sugiyama, Krauledat, and
  M{\~A}{\v{z}}ller}{Sugiyama et~al.}{2007}]{sugiyama2007covariate}
Sugiyama, M., Krauledat, M., and M{\~A}{\v{z}}ller, K.-R. (2007).
\newblock Covariate shift adaptation by importance weighted cross validation.
\newblock {\em Journal of Machine Learning Research} {\bf 8,} 985--1005.

\bibitem[\protect\citeauthoryear{Tipton}{Tipton}{2013}]{tipton2013improving}
Tipton, E. (2013).
\newblock Improving generalizations from experiments using propensity score
  subclassification: Assumptions, properties, and contexts.
\newblock {\em Journal of Educational and Behavioral Statistics} {\bf 38,}
  239--266.

\bibitem[\protect\citeauthoryear{Wang and Zubizarreta}{Wang and
  Zubizarreta}{2020}]{wang2020minimal}
Wang, Y. and Zubizarreta, J.~R. (2020).
\newblock Minimal dispersion approximately balancing weights: asymptotic
  properties and practical considerations.
\newblock {\em Biometrika} {\bf 107,} 93--105.

\bibitem[\protect\citeauthoryear{Westreich, Edwards, Lesko, Stuart, and
  Cole}{Westreich et~al.}{2017}]{westreich2017transportability}
Westreich, D., Edwards, J.~K., Lesko, C.~R., Stuart, E., and Cole, S.~R.
  (2017).
\newblock Transportability of trial results using inverse odds of sampling
  weights.
\newblock {\em American journal of epidemiology} {\bf 186,} 1010--1014.

\bibitem[\protect\citeauthoryear{Wong and Chan}{Wong and
  Chan}{2018}]{wong2018kernel}
Wong, R.~K. and Chan, K. C.~G. (2018).
\newblock Kernel-based covariate functional balancing for observational
  studies.
\newblock {\em Biometrika} {\bf 105,} 199--213.

\bibitem[\protect\citeauthoryear{Yang, Kim, and Song}{Yang
  et~al.}{2020}]{yang2020doubly}
Yang, S., Kim, J.~K., and Song, R. (2020).
\newblock Doubly robust inference when combining probability and
  non-probability samples with high dimensional data.
\newblock {\em Journal of the Royal Statistical Society: Series B (Statistical
  Methodology)} {\bf 82,} 445--465.

\bibitem[\protect\citeauthoryear{Zhao and Percival}{Zhao and
  Percival}{2016}]{zhao2016entropy}
Zhao, Q. and Percival, D. (2016).
\newblock Entropy balancing is doubly robust.
\newblock {\em Journal of Causal Inference} {\bf 5,} 20160010.

\end{thebibliography}
\endgroup

\section*{Supporting Information}
Web Appendix with technical proofs referenced in Section \ref{sec:theory} and additional simulations
referenced in Section \ref{sec:simu}  as well as the software for the proposed method in R are available with this paper at the Biometrics website on Wiley Online Library.

\end{document}